\documentclass[twocolumn,tighten]{aastex62}
\usepackage{amsfonts,amsmath,amssymb}
\usepackage{txfonts}
\usepackage{graphicx}
\usepackage{bm}
\usepackage{verbatim}
\usepackage{hyperref}
\usepackage{xcolor}
\usepackage{units}
\usepackage{amsbsy}
\usepackage{esint}

\newcommand{\pop}{Phys. Plasmas}

\newcommand{\jfm}{J. Fluid Mach.}

\newcommand{\apjss}{Astrophys. J. Supp.}

\newcommand{\jplp}{J. Plasma Phys.}
\newcommand{\nature}{Nature}

\newcommand{\pss}{Plan. Space Sci.}
\newcommand{\pr}{Phys. Rev.}
\newcommand{\arfm}{Ann. Rev. Fluid Mech.}

\newcommand{\arcm}{Ann. Rev. Cond. Matt.}

\newcommand{\dansssr}{Dok. Akad. Nauk SSSR}

\newcommand{\driftvel}{{\bf w}_{s}}
\newcommand{\driftvelmag}{{w}_{s}}
\newcommand{\driftvelhat}{\hat{{\bf w}}_{s}}

\begin{document}
\title{Resonant Drag Instability of Grains Streaming in Fluids}

\author[0000-0001-8479-962X]{J.~Squire}
\correspondingauthor{J.~Squire}
\email{jsquire@caltech.edu}
\author[0000-0003-3729-1684]{P.~F.~Hopkins}
\affiliation{Theoretical Astrophysics, 350-17, California Institute of Technology, Pasadena, CA 91125, USA}
\affiliation{Walter Burke Institute for Theoretical Physics, Pasadena, CA 91125, USA}


\begin{abstract}
We show that grains streaming through a fluid are generically unstable if their
velocity, projected along some direction, matches the phase velocity of a fluid wave (linear oscillation).  This
can occur whenever grains stream faster than any fluid wave. The wave itself can be quite general---sound waves, 
magnetosonic waves,  epicyclic oscillations, and Brunt-V\"ais\"al\"a oscillations each generate instabilities, for example. We derive a simple expression for the growth rates of these ``resonant drag instabilities'' (RDI). This expression (i)
illustrates why such instabilities are so virulent and generic, and (ii) allows for simple analytic computation
of RDI growth rates and properties for different fluids. 
As examples, we introduce several new instabilities, 
which could see application across a variety of physical systems from atmospheres to protoplanetary disks, the interstellar medium, and galactic outflows.
The matrix-based resonance formalism we introduce can also be applied more generally in other (nonfluid) contexts, providing a simple means for calculating and understanding the stability properties of interacting systems. 
\end{abstract}

\section{Introduction}
Many astrophysical fluids---e.g., the interstellar medium \citep{2001RvMP...73.1031F}, disks \citep{2011ARA&A..49..195A}, our solar system \citep{2015ApJ...812..139K}---are laden with solid grains, or dust. Because they contain a large proportion of the available metals,  dust is   fundamental  to key astrophysical processes such as star and planet formation (see, e.g., \citealt{1996rdfs.conf.dust.star.formation,2004tcu..conf..213D,chiang:2010.planetesimal.formation.review}). It is thus crucial to  understand  dust-fluid dynamics \citep{saffman_1962,doi:10.1146/annurev-conmatphys-031115-011538}: how the phases interact through drag and/or Lorentz forces;  
what forces influence the fluid  but not  dust grains, and vice versa.

In this Letter, we ask a simple question: if dust grains  stream through a fluid (gas) with some constant relative velocity $\driftvel$, is the coupled system stable?   We show, quite generally, that this system is usually unstable if the 
phase speed of a wave in the fluid  matches the projection of $\driftvel$  along the wave propagation direction. This resonant fluid wave is stationary in the frame of the dust grains and couples very efficiently to grain density perturbations. This
usually renders the wave unstable because  it can  feed off the energy in the background drift, causing clumping of 
the grains in space as the instability grows. Many fluid waves---e.g., sound waves, magnetosonic 
waves, Brunt-V\"ais\"al\"a oscillations, or epicyclic oscillations---can cause such a  ``resonant drag instability'' (RDI). 
Further, because the fluid wave can be destabilized 
at an angle to the grain's velocity $\driftvel$, any streaming motion \emph{faster} than the phase speed can cause an RDI.
For example, in hydrodynamics, the RDI occurs whenever $\driftvelmag=|\driftvel|>c_{s}$ (the sound speed), while in 
magnetohydrodynamics (MHD) or a stratified fluid,  
the RDI is possible for \emph{any} $\driftvelmag$.


A relative dust-to-gas streaming velocity $\driftvel$ can occur for a variety of reasons. In many astrophysical systems---e.g., near active galactic nucleii \citep{krolik:clumpy.torii,thompson:rad.pressure,hopkins:torus}, in the envelopes of cool stars \citep{1989A&A...223..227D,2012Natur.484..220N}, or near star-forming regions \citep{franco:dust.rad.pressure.galactic.fountain,2005ApJ...618..569M}---radiation pressure  more strongly affects the dust grains than the gas. As  grains are accelerated, they drag the gas with them, reaching a
terminal $\driftvel$ when the drag force nearly balances the radiative force  \citep{1972ApJ...178..423G,1993ApJ...410..701N}.
Another source of relative drift occurs when the gas---but not the dust---is supported by thermal pressure against gravity. In, for example, planetary atmospheres or astrophysical disks, this causes grains to settle in the direction of gravity \citep{1973ApJ...183.1051G,1986Icar...67..375N}. However, despite these diverse mechanisms that cause a relative drift, in each case, the stability of the coupled dust gas system
can be calculated in the frame where the gas is stationary (a bulk velocity or linear acceleration does not change the system's spectral stability properties; \citealt{Hopkins:2017rdi}).
Thus, in this Letter we simply \emph{prescribe} $\driftvel$,  remaining agnostic about its origin. We also  assume a homogenous background gas and dust density (the local approximation), neglect dissipative processes (e.g., viscosity) in the gas, and assume grains interact with the gas only through drag forces (neglecting, e.g., grain charge and dusty plasma effects; \citealt{2001ApJS..134..263W,RAO1990543,2001PhPl....8.1791S,2004tcu..conf..213D}).  Detailed physical applications are treated in companion papers \citep{Hopkins:2017rdi,Squire:2017rdi,Hopkins:2018mhdrdi}; the purpose of this letter is to introduce the basic  mathematical formalism and structure of dust-gas
RDIs.

Following a general derivation of the RDI, this Letter is organized into three examples; hydrodynamics, MHD, and  stratified fluids. The general nature of these 
instabilities has not (to our knowledge) been discussed in previous works, although specific manifestations of the hydrodynamic instability are
studied in \citet{1993ESOC...46...60M,mastrodemos:dust.gas.coupling.cool.winds.spherical.symmetry,1997IAUS..180..151D}, and  instabilities of a streaming neutral gas in MHD  are  treated in detail  in \citet{2002MNRAS.337..117T}. 
We also note that the widely studied ``streaming instability'' of grains in  protoplanetary disks \citep{2000Icar..148..537G,2005ApJ...620..459Y,2007Natur.448.1022J}, is an RDI with disk epicyclic
 oscillations, although its resonant nature has not (to our knowledge) been recognized previously.
Similar ideas are more generally related to a variety of  instabilities in fluids and plasmas (e.g., \citealt{1967JPlPh...1...75K,Childress:1975,doi:10.1146/annurev.fluid.35.101101.161151,Verscharen:2013b}).
Throughout this letter, we  study  the RDI only exactly at resonance, although each example  also displays an array of other slower-growing instabilities (see \citealt{Hopkins:2017rdi}; as shown below, resonant modes are always the fastest growing at low grain concentrations).

\vspace{-0.5cm}
\section{Basic theory of resonance instability}

Before deriving  the RDI dispersion relation for various specific fluid systems, 
we consider the mathematics of interacting linear systems.   
Our purpose here is twofold: first, these  results show why resonances generically
  lead to  virulent instabilities; second, we will derive 
 formulae for the RDI growth rate  in terms of fluid eigenmodes [Eqs.~\eqref{eq: general eval pert} and \eqref{eq: general eval pert 3}]. These
formulae allow  the dispersion relation of different RDIs to be calculated with 
relative ease,  
even for complicated fluid systems 
 (e.g., MHD in 3-D). Aspects of these results  are related to ``Krein collisions'' in the theory of Hamiltonian mechanics \citep{Krein,kirillov2013nonconservative}, although we do not restrict ourselves to Hamiltonian systems.

Consider an arbitrary system of equations that describes the motion of a coupled system of fluid, denoted $\bm{f}$ [e.g., with density and velocity variables, $\bm{f} = (\rho,\bm{u},\dots)$], and dust, $\bm{a} = (\rho_{d},\bm{v})$ (the dust continuum density and velocity). For small perturbations ($\bm{f}=\langle\bm{f}\rangle+\delta\bm{f},\,\bm{a}=\langle\bm{a}\rangle+\delta\bm{a}$), which are Fourier decomposed in space and time ($\delta g(\bm{x},t)=\delta ge^{i\bm{k}\cdot\bm{x}-i\omega t}$), the linearized equations of motion [see Eqs.~\eqref{eq: general dust system 1}--\eqref{eq: general dust system 2} below] take the form of a generic eigenvalue problem,
\begin{gather}\omega\left(\begin{array}{c}\delta\bm{a}\\\delta\bm{f}\end{array}\right)
=\mathbb{T}\left(\begin{array}{c}\delta\bm{a}\\\delta\bm{f}\end{array}\right)
=\left(\mathbb{T}_{0}+\mu\,\mathbb{T}^{(1)}\right)\left(\begin{array}{c}\delta\bm{a}\\\delta\bm{f}\end{array}\right),\nonumber\\
\mathbb{T}_{0}=\left(\begin{array}{cc}\mathcal{A}&\mathcal{C}\\0&\mathcal{F}\end{array}\right),
\quad\mathbb{T}^{(1)}=\left(\begin{array}{cc}\mathcal{T}^{(1)}_{AA}&\mathcal{T}^{(1)}_{AF}\\\mathcal{T}^{(1)}_{FA}&\mathcal{T}^{(1)}_{FF}\end{array}\right).\label{eq: general linear}\end{gather}
Here $\mathbb{T}\equiv\mathbb{T}_{0}+\mu\mathbb{T}^{(1)}$ is the full linearized system of equations, decomposed (without loss of generality) into the block-matrix form $\mathbb{T}_{0}$ (composed of $\mathcal{A},\,\mathcal{F},\,\mathcal{C}$) and $\mu\mathbb{T}^{(1)}$ (where $\mu\equiv\rho_{d}/\rho$ is the ratio of dust to fluid continuum densities). Submatrix $\mathcal{F}$ describes the fluid in the absence of dust, $\mathcal{A}$ describes  dust in the absence of fluid motions,  $\mathcal{C}$ couples the dust to the fluid (e.g., drag on the dust), and $\mu\mathbb{T}^{(1)}$ contains any coupling of the fluid to the dust (e.g., the back-reaction from dust, in $\mathcal{T}^{(1)}_{FA}$). If $\mathrm{Im}(\omega)>0$, the system is unstable (perturbations grow).


Now stipulate that $\mathcal{A}$ and $\mathcal{F}$ share an eigenvalue, $\omega=\omega_{0}$, which we define as a \emph{resonance}. 
It is most instructive to examine the limit $\mu\ll1$; i.e., to ask what happens to the eigenvalue $\omega_{0}$ as the dust starts influencing the fluid's dynamics. 
Mathematically, this is the eigenvalue perturbation, $\omega=\omega_{0}+\omega^{(1)}+\dots$, due to $\mu\mathbb{T}^{(1)}$. Assuming $\omega_{0}$ is a semisimple eigenvalue of $\mathcal{A}$ and $\mathcal{F}$ individually,  define its right and left eigenvectors,
\begin{equation}(\mathcal{A}-\omega_{0}\mathbb{I})\xi_{\mathcal{A}}^{R}=0\quad\text{and}\quad\xi_{\mathcal{A}}^{L}(\mathcal{A}-\omega_{0}\mathbb{I})=0,\end{equation}
with $\xi_{\mathcal{A}}^{L}\xi_{\mathcal{A}}^{R}=1$, $\mathbb{I}$ the identity matrix, and
 equivalent definitions  for $\mathcal{F}$ with $\xi^{L,R}_{\mathcal{F}}$.  
Using the block structure of $\mathbb{T}_{0}$ \citep{dobson2001strong}, one  can  show that if $\xi_{\mathcal{A}}^{L}\mathcal{C}\xi_{\mathcal{F}}^{R} \neq 0$, then $\omega_{0}$ is a \emph{defective eigenvalue} of $\mathbb{T}_{0}$,
meaning it has only one associated eigenvector.
This implies that  standard perturbation theory ($\omega^{(1)}=\mu\,\xi_{\mathbb{T}_{0}}^{L}\mathbb{T}^{(1)}\xi_{\mathbb{T}_{0}}^{R}$, as  familiar from quantum mechanics) does not apply. Instead, we  perturb the  Jordan block,
\begin{equation}J=\left(\begin{array}{cc}\omega_{0}&1\\0&\omega_{0}\end{array}\right)=(p^{L}p^{R})^{-1}p^{L}\mathbb{T}_{0}p^{R},\label{eq: 2*2 jordan block}\end{equation}
where $p^{R}$ ($p^{L}$) is the right (left) generalized eigenvector block for $\omega_{0}$. 
Importantly, defective eigenvalues are particularly sensitive to perturbations \citep{katoBook}: if $\omega_{0}$ corresponds to an $n\times n$ Jordan block,
then (in general) $\omega^{(1)}\sim\mathcal{O}(\mu^{1/n})$ \citep{moro2002first}.
Thus, the $\mathcal{O}(\mu)$ perturbation to $\mathbb{T}_{0}$ causes an $\mathcal{O}(\mu^{1/2})$ perturbation to $\omega$.

We compute $\omega^{(1)}$ by calculating the eigenvalues of $\mathbb{T}_{0}+\mu\,\mathbb{T}^{(1)}$ in the generalized eigenvector basis \eqref{eq: 2*2 jordan block}.
The result is particularly simple and useful:
\begin{equation}\omega=\omega_{0}\pm\mu^{1/2}\left[\left(\xi_{\mathcal{F}}^{L}\,\mathcal{T}^{(1)}_{FA}\,\xi_{\mathcal{A}}^{R}\right)\left(\xi_{\mathcal{A}}^{L}\,\mathcal{C}\,\xi_{\mathcal{F}}^{R}\right)\right]^{1/2}+\dots,\label{eq: general eval pert}\end{equation}
which depends  on the coupling terms  ($\mathcal{C}$, coupling $\bm{a}$ to $\bm{f}$, and $\mathcal{T}^{(1)}_{FA}$, coupling $\bm{f}$ to $\bm{a}$)
only through simple matrix multiplication.
Noting that $\xi^{L,R}_{\mathcal{F}},\,\xi^{L,R}_{\mathcal{A}},\,\mathcal{C}$, and $\mathcal{T}_{FA}^{(1)}$ are in general complex, Eq.~\eqref{eq: general eval pert} reveals why resonance instabilities are so virulent and so generic: the perturbation causes an instability [$\mathrm{Im}(\omega)>0$] unless $(\xi_{\mathcal{F}}^{L}\,\mathcal{T}^{(1)}_{FA}\,\xi_{\mathcal{A}}^{R}) (\xi_{\mathcal{A}}^{L}\,\mathcal{ C}\,\xi_{\mathcal{F}}^{R})$ is real and positive (or zero). Moreover, for $\mu\ll1$, such modes grow more rapidly [$\mathrm{Im}(\omega)\sim\mathcal{O}(\mu^{1/2})$]  than the usual perturbation theory expectation [$\mathrm{Im}(\omega)\sim\mathcal{O}(\mu)$].

At  short-wavelengths, the dust operator $\mathcal{A}$ itself becomes defective in $\omega_{0}$ [see Eq.~\eqref{eq: dust matrix form}], and we must generalize Eq.~\eqref{eq: general eval pert} to 3 blocks:
\begin{equation}\mathbb{T}_{0}=\left(\begin{array}{ccc}\mathcal{A}_{1}&\mathcal{C}_{12}&0\\0&\mathcal{A}_{2}&\mathcal{C}_{2F}\\0&0&\mathcal{F}\end{array}\right),\label{eq: general 3*3 system}\end{equation}
where $\mathcal{A}_{1}$, $\mathcal{A}_{2}$, and $\mathcal{F}$ share an eigenvalue $\omega_{0}$.
One obtains,
\begin{equation}\omega^{(1)}=s_{3}\,\mu^{1/3}\left[\left(\xi_{\mathcal{F}}^{L}\,\mathcal{T}^{(1)}_{F1}\,\xi_{\mathcal{A}_{1}}^{R}\right)\left(\xi_{\mathcal{A}_{1}}^{L}\,\mathcal{C}_{12}\,\xi_{\mathcal{A}_{2}}^{R}\right)\left(\xi_{\mathcal{A}_{2}}^{L}\,\mathcal{C}_{2F}\,\xi_{\mathcal{F}}^{R}\right)\right]^{1/3}+\dots,\label{eq: general eval pert 3}\end{equation}
where  $\mathcal{T}^{(1)}_{F1}$ is lower left block of  $\mathbb{T}^{(1)}$ and the values of $s_{3}=(1,-1/2\pm i\sqrt{3}/2)$ solve $s_{3}^{3}=1$. The perturbed system is \emph{always} unstable for one $s_{3}$ unless Eq.~\eqref{eq: general eval pert 3} is zero.
 
\section{Dust-gas systems}

We now specify $\mathcal{A}$ in Eq.~\eqref{eq: general linear}, modeling the grains as a pressureless fluid \citep{doi:10.1146/annurev.fl.15.010183.001401,Jacquet:2011cy}, interacting with the gas fluid through a generic neutral drag force. The 
formalism is easily extended to incorporate more complex dust and drag physics (e.g.,  grain charge; \citealp{Hopkins:2018mhdrdi}).
We keep the fluid system (i.e., the $\mathcal{F}$ matrix) general at this stage, but assume it has density and velocity variables $\rho$ and $\bm{u}$  (in addition to other properties, e.g., magnetic field). We work in  the frame where the fluid is stationary (which may have constant linear acceleration; \citealp{Hopkins:2017rdi}), with the grains streaming at velocity $\driftvel=\driftvelmag\driftvelhat$. 

On a homogenous background (with $\langle\cdot\rangle$  denoting a spatial average), the linearized and Fourier-decomposed continuum dust density, $\rho_{d}=\mu\langle\rho\rangle(1+\delta\rho_{d})$, and velocity, $\bm{v}=\langle\bm{v}\rangle+\delta\bm{v}=\driftvel+\delta\bm{v}$, satisfy
\begin{gather}(-i\omega+i\driftvel\cdot\bm{k})\delta\rho_{d}+i\bm{k}\cdot\delta\bm{v}=0,\label{eq: general dust system 1}\\
(-i\omega+i\driftvel\cdot\bm{k})\delta\bm{v}=-\delta{\bf F}_{\mathrm{drag}}(\driftvelmag,\bm{u},\rho,\bm{v}).\label{eq: general dust system 2}\end{gather}
Here $\delta{\bf F}_{\mathrm{drag}}$ is the linearized drag acceleration, which we take as
${\bf F}_{\mathrm{drag}}=({\bm{v}-\bm{u}})/{t_{s}}$
where $t_{s}(\rho,|\bm{u}-\bm{v}|)$ is the ``stopping time.'' We   parameterize $t_{s}$
through 
${\delta t_{s}}/{\langle t_{s}\rangle}=-\zeta_{s}\,{\delta\rho}/{\langle\rho\rangle}-\zeta_{w}{\driftvelhat\cdot(\delta\bm{v}-\delta\bm{u})}/\driftvelmag$, 
where $\langle t_{s}\rangle=t_{s}(\langle\rho\rangle,\driftvelmag)$.
This form of the dust-fluid drag, determined by $\zeta_{s}$ and $\zeta_{w}$, encompasses many drag laws for uncharged grains in a polytropic fluid. For example, when the grain size $R_{d}$ is smaller than the gas mean free path  $\lambda_{\mathrm{mfp}}$ (``Epstein drag''; \citealp{Epstein}), 
\begin{equation}t_{s}\approx\frac{a_{\gamma}^{1/2}m_{d}}{\pi\rho c_{s}R_{d}^{2}}\left(1+a_{\gamma}\frac{|\bm{v}-\bm{u}|^{2}}{c_{s}^{2}}\right)^{-1/2},\quad a_{\gamma}\equiv\frac{9\pi\gamma}{128},\label{eq: epstein drag}\end{equation}
 which gives $\zeta_{s}^{\mathrm{Ep}}=(\gamma+1+2\,\tilde{a}_{E})/(2+2\,\tilde{a}_{E})$, $\zeta_{w}^{\mathrm{Ep}}=\tilde{a}_{E}/(1+\tilde{a}_{E})$ (here $\tilde{a}_{E}\equiv a_{\gamma}\,(\driftvelmag/c_{s})^{2}$, $m_{d}$ is the mass of individual grains, and $\gamma$ is the fluid polytropic index). 
 The coefficients $\zeta_{s}$ and $\zeta_{w}$ for other drag laws (e.g., Stokes or Coulomb drag; \citealp{1979ApJ...231...77D}) can be  calculated in a similar manner \citep{Hopkins:2017rdi}. 
From momentum conservation, the drag on the {fluid} (contained in $\mathbb{T}^{(1)}$) is $+(\rho_{d}/\rho){\bf F}_{\mathrm{drag}}$.

With $\delta\bm{a}=(\delta\rho_{d},\delta\bm{v}$), Eqs.~\eqref{eq: general dust system 1}--\eqref{eq: general dust system 2} give
\begin{equation}\mathcal{A}=\left(\begin{array}{cc}\omega_{0}&\bm{k}^{T}\\0&\omega_{0}\mathbb{I}+\mathcal{D}_{\mathrm{drag}},\end{array}\right),
\quad\mathcal{C}=\left(\begin{array}{c}0\\\mathcal{C}_{\bm{v}},\end{array}\right),\label{eq: dust matrix form}\end{equation}
where $\omega_{0}=\bm{k}\cdot\driftvel=k\driftvelmag\psi_{kw}$ is the resonant eigenmode (we define $\bm{k}=k\hat{\bm{k}}$ and $\psi_{kw}\equiv\hat{\bm{k}}\cdot\driftvelhat$  for convenience), $\mathcal{D}_{\mathrm{drag}}= -i\,(\mathbb{I}+\zeta_{w} \driftvelhat\driftvelhat^{T})/\langle t_{s}\rangle$, and $\mathcal{C}_{\bm{v}}$ follows  from the drag law [e.g., if $\delta\bm{f} = (\delta\rho/\langle\rho\rangle,\delta\bm{u})$, $\mathcal{C}_{\bm{v}} = i\,(-\zeta_{s}\,\driftvel,\,\mathbb{I}+\zeta_{w}\,\driftvelhat\driftvelhat^{T})/\langle t_{s}\rangle$].  
Evaluating Eq.~\eqref{eq: general eval pert}, we derive the RDI growth rate, which is valid when $\omega_{0}$ is also 
an eigenvalue of $\mathcal{F}$,
\begin{equation}\omega=\omega_{0}\pm i\mu^{1/2}\left[(\xi_{\mathcal{F}}^{L}\mathcal{T}^{(1)}_{\rho_{d}})\,(\bm{k}^{T}\mathcal{D}_{\mathrm{drag}}^{-1}\mathcal{C}_{\bm{v}}\xi_{\mathcal{F}}^{R})\right]^{1/2}+\dots,\label{eq: dust eval pert}\end{equation}
where $\mathcal{T}^{(1)}_{\rho_{d}}=\mathcal{T}^{(1)}_{FA}\xi^{R}_{\mathcal{A}}=i\driftvel/\langle t_{s}\rangle$ is the left column of $\mathcal{T}^{(1)}_{FA}$.  

As $k$ increases, Eq.~\eqref{eq: dust eval pert} becomes invalid because $\mathcal{A}$ is nearly defective in $\omega_{0}$ when $\bm{k}^{T}$ dominates over 
$\mathcal{D}_{\mathrm{drag}}$. The  theory is then modified to the triply defective case \eqref{eq: general eval pert 3}, which treats both $\mathcal{D}_{\mathrm{drag}}$ and  $\mu\,\mathbb{T}^{(1)}$ as perturbations.
Using Eq.~\eqref{eq: general eval pert 3}  with $\mathcal{A}_{1}=\omega_{0}$, $\mathcal{A}_{2}=\omega_{0}\mathbb{I}$, $\mathcal{C}_{12}=\bm{k}^{T}$, $\xi_{\mathcal{A}_{2}}^{R}=\hat{\bm{k}}$, and
$\mathcal{C}_{2F}=\mathcal{C}_{\bm{v}}$, 
one obtains,
\begin{equation}\omega=\omega_{0}+s_{3}\,\mu^{1/3}\left[(\xi_{\mathcal{F}}^{L}\mathcal{T}^{(1)}_{\rho_{d}})\,(\bm{k}^{T}\mathcal{C}_{\bm{v}}\xi_{\mathcal{F}}^{R})\right]^{1/3}+\dots\label{eq: dust eval pert 3}\end{equation}
for the  ``high-$k$'' RDI.
From the characteristic polynomial of Eq.~\eqref{eq: general 3*3 system}, one finds that the transition between Eqs.~\eqref{eq: dust eval pert 3}  and \eqref{eq: dust eval pert} occurs when the two are
approximately equal, at $\mu k\sim(\xi_{\mathcal{F}}^{L}\mathcal{T}^{(1)}_{\rho_{d}})^{-1}(\hat{\bm{k}}^{T}\mathcal{C}_{\bm{v}}\xi_{\mathcal{F}}^{R})^{2}(\hat{\bm{k}}^{T}\mathcal{D}_{\mathrm{drag}}^{-1}\mathcal{C}_{\bm{v}}\xi_{\mathcal{F}}^{R})^{-3}$. 

Finally, we note that the result \eqref{eq: dust eval pert} is also not valid when $\omega_{0}\ll\mu\driftvelmag\langle t_{s}\rangle$ (i.e., when $\mu\mathbb{T}^{(1)}$ is larger than $\omega_{0}$), although instabilities generically persist in this regime  \citep{Hopkins:2017rdi}.
Inaccuracies can also
arise near certain special points---e.g., when $\hat{\bm{k}}^{T}\mathcal{D}_{\mathrm{drag}}^{-1}\mathcal{C}_{\bm{v}}\xi_{\mathcal{F}}^{R}\approx0$---if the ordering used to derive Eqs.~\eqref{eq: dust eval pert} and \eqref{eq: dust eval pert 3} becomes inaccurate. A small background dust pressure $\langle P_{d}\rangle$ causes the dust eigenmode to be weakly damped, $\mathrm{Im}(\omega_{0,\mathrm{dust}})\sim-\langle P_{d}/\rho_{d}\rangle k^{2}\langle t_{s}\rangle$,  and our results are valid for $ \omega^{(1)}\gg|\mathrm{Im}(\omega_{0,\mathrm{dust}})|$.

 \begin{figure}
\begin{center}
\includegraphics[width=1.0\columnwidth]{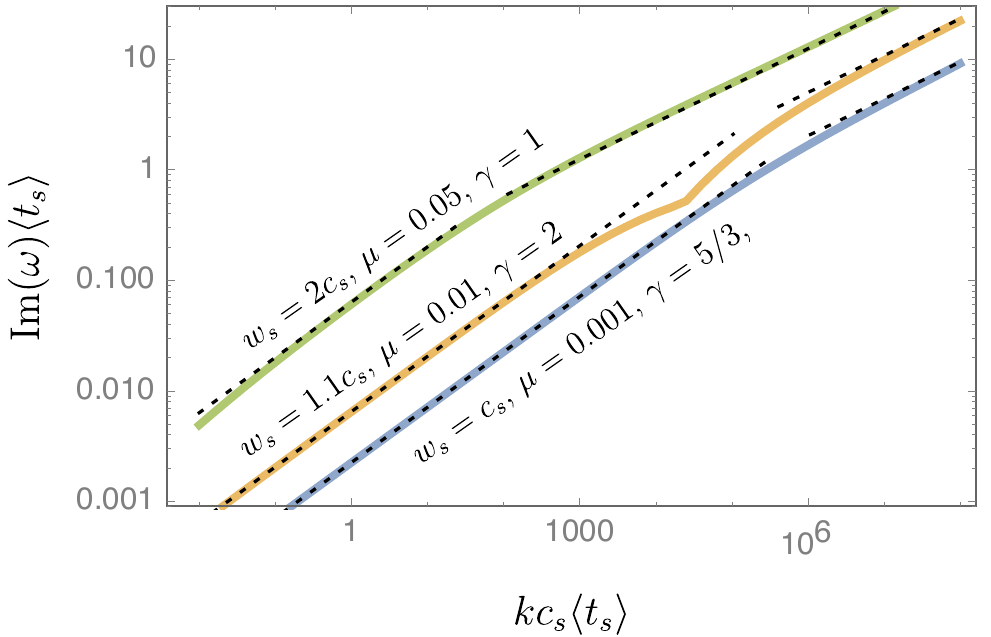}
\caption{\emph{Acoustic RDI in hydrodynamics:} resonant mode growth rate $\mathrm{Im}[\omega(k)]$, for several streaming velocities $\driftvelmag/c_{s}$ with  $\psi_{kw}=\hat{\bm{k}}\cdot\driftvelhat=c_{s}/\driftvelmag$ (i.e., the resonant $\hat{\bm{k}}$). Thick colored lines show numerical solutions of the exact dispersion relation of the full coupled dust-gas system (Eqs.~\eqref{eq: general dust system 1}--\eqref{eq: general dust system 2}, coupled to the compressible Euler equations), and  black dotted lines show  analytic expressions \eqref{eq: hydro mid k}--\eqref{eq: hydro high k}. We use a variety of parameters as labeled, and 
 Epstein drag \eqref{eq: epstein drag} for $\zeta_{s}$ and $\zeta_{w}$. The dip around 
$kc_{s}\langle t_{s}\rangle\sim10^{5}$ for $\driftvelmag=1.1c_{s}$ occurs because the parameters ($\zeta_{s}=1.33,\,\zeta_{w}=0.35$) lie near $\zeta_{s}=1+\zeta_{w}$ (see Eq.~\eqref{eq: hydro mid k} and \citealt{Hopkins:2017rdi}).}
\label{fig: hydro}
\end{center}
\end{figure}

\section{Examples}
\subsection{Neutral hydrodynamics}
We now consider the RDI  in a variety of 
physical fluids (prescribing $\mathcal{F}$), starting with sound waves in compressible hydrodynamics. This amounts to setting $\mathcal{F}$ to describe a neutral compressible gas.
 This instability is explored in detail, including discussion of mode structure and astrophysical applications, in \citet{Hopkins:2017rdi}. Noting the symmetry of the problem, we set  $\driftvelhat=\hat{\bm{z}}$ and
consider  2-D perturbations ($\bm{k}=k_{x}\hat{\bm{x}}+k_{z}\hat{\bm{z}}$). The linearized sound-wave eigenmodes for $(\delta \rho/\langle\rho\rangle,\delta u_{x},\delta u_{z})$ are  
$\xi^{R}_{\mathcal{F}\pm}=2^{-1/2}(\pm c_{s}^{-1},k_{x}/k,k_{z}/k)^{T},$ $\xi^{L}_{\mathcal{F}\pm}=2^{-1/2}(\pm c_{s},k_{x}/k,k_{z}/k)$, with eigenvalues
  $\omega_{\pm}=\pm kc_{s}$.
We see that for
  for $\driftvelmag>c_{s}$ there is \emph{always a resonant mode}---propagating in the direction $\psi_{kw}=k_{z}/k=c_{s}/\driftvelmag$---for which $\omega_{0}=\bm{k}\cdot\driftvel=kc_{s}$ for all $k$. The RDI growth rate thus {increases indefinitely as} $k\rightarrow\infty$  (neglecting viscosity, which damps the RDI once $k\gtrsim\lambda_{\mathrm{mfp}}^{-1}$).

Evaluating  Eq.~\eqref{eq: dust eval pert}, we obtain an approximate expression (to leading order in matrix perturbation theory) for this ``acoustic RDI,'' 
\begin{equation}\omega\approx kc_{s}+s_{2}\,\mu^{1/2}k^{1/2}\left[{\frac{c_{s}}{2\langle t_{s}\rangle}\left(1-\frac{\zeta_{s}}{1+\zeta_{w}}\right)}\right]^{1/2},\label{eq: hydro mid k}\end{equation}
where $s_{2}=\pm(1+i)/\sqrt{2}$ solves $s_{2}^{2}=i$.
For very high-frequency modes, Eq.~\eqref{eq: dust eval pert 3} gives
\begin{equation}\omega\approx kc_{s}+s_{3}\,\mu^{1/3}k^{1/3}\left[\frac{c_{s}}{2\langle t_{s}\rangle^{2}}\,\left(\zeta_{s}-1-\psi_{kw}^{2}\zeta_{w}\right)\,\right]^{1/3}.\label{eq: hydro high k}\end{equation}
In Fig.~\ref{fig: hydro}, we show several examples, comparing Eqs.~\eqref{eq: hydro mid k}--\eqref{eq: hydro high k} with direct numerical solutions of the exact linearized grain-fluid dispersion relation for neutral, inviscid hydrodynamics and pressure-free grains coupled via Epstein drag. This confirms the instabilities exist, and shows that our analytic expressions are accurate where they apply. While the analytic Eqs.~\eqref{eq: hydro mid k}--\eqref{eq: hydro high k} are valid only at $k_{z}/k=c_{s}/\driftvelmag$, the system is also unstable at other mode angles and wavenumbers, albeit with lower growth rates ($\mathrm{Im}(\omega)\sim\mathcal{O}(\mu)$ when $\mu\ll1$;  \citealt{Hopkins:2017rdi}).

 \begin{figure}
\begin{center}
\includegraphics[width=1.0\columnwidth]{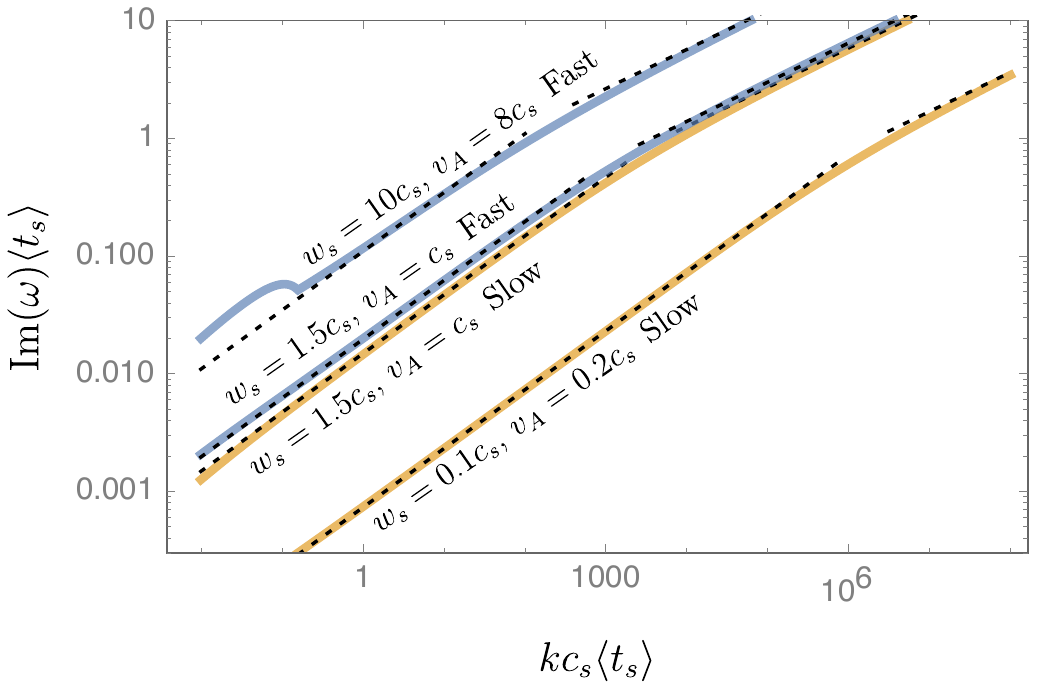}
\caption{\emph{Magnetosonic RDI in MHD:} blue (orange) lines show numerically calculated RDI growth rates for grains resonant with the 
fast (slow) mode with streaming angle $\theta_{w}=45^{\circ}$ and $\mu=0.01$ ($\mu=0.05$ for the $\driftvelmag=10c_{s}$ fast wave). Dotted lines  show the analytic predictions \eqref{eq: MHD mid k} [Eq.~\eqref{eq: dust eval pert 3} at high $k$].  In each case we calculate $\zeta_{s}$ and $\zeta_{w}$ assuming Epstein drag \eqref{eq: epstein drag} with $\gamma=5/3$. The resonant mode directions, $\hat{\bm{k}}=(\cos\phi\sin\theta,\sin\phi\sin\theta,\cos\theta)$, 
are: $\theta=70^{\circ},\,\phi=49.6^{\circ}$ (slow wave, $\driftvelmag=0.1c_{s}$);  $\theta=40^{\circ},\,\phi=108.4^{\circ}$ (slow wave, $\driftvelmag=1.5c_{s}$); $\theta=20^{\circ},\,\phi=63.5^{\circ}$ (fast wave, $\driftvelmag=1.5c_{s}$); $\theta=30^{\circ},\,\phi=57.6^{\circ}$ (fast wave, $\driftvelmag=10c_{s}$). 
The low-$k$ discrepancy of the fast-wave prediction at $\driftvelmag=10c_{s}$
is due to an additional instability.}
\label{fig: MHD}
\end{center}
\end{figure}

\subsection{Magnetohydrodynamics}
With more waves (Alfv\'en, slow, and fast modes), specifying $\mathcal{F}$ to describe MHD (including a magnetic field $\bm{B}$ in $\bm{f}$) allows for richer resonance 
phenomena. This structure, including the effects of grain charge (e.g., Lorentz forces on grains), is explored in detail in \citet{Hopkins:2018mhdrdi},  along with discussion of the diverse array of astrophysical environments where MHD RDIs could be important.
As in hydrodynamics, MHD waves have constant phase velocities (for a given $\hat{\bm{k}}$), 
and the growth rate of the RDI increases indefinitely as $k\rightarrow\infty$. 
The resonant condition is $\driftvelmag\,\psi_{kw}=V_{\mathrm{wave}}(\hat{\bm{k}})$ (where $V_{\mathrm{wave}}$ is the wave
phase velocity), and we take $\driftvelhat=\sin\theta_{w}\hat{\bm{x}}+\cos\theta_{w}\hat{\bm{z}}$ with  $\bm{B}=B_{0}\hat{\bm{z}}$.
Remarkably, because the slow mode phase velocity approaches zero as $\hat{\bm{k}}\cdot \bm{B}\rightarrow0$, an instability occurs---with $\mathrm{Im}(\omega)\rightarrow\infty$ as $k\rightarrow\infty$---for \emph{any} $\driftvelmag$, so long as $\theta_{w}\neq0$.

Evaluating~\eqref{eq: dust eval pert}, we find that  Alfv\'en waves do not cause 
a mid-$k$ RDI (the product in square brackets is zero for neutral grains), while resonance with slow or 
fast waves triggers the ``magnetosonic RDI,''
\begin{equation}\omega_{\pm}\approx kv_{\pm}c_{s}+s_{2}\left(\frac{k\mu c_{s}}{\langle t_{s}\rangle}\right)^{1/2}\left[\frac{\tilde{\zeta_{w}}-\zeta_{s}}{\tilde{\zeta_{w}}}\left(\frac{v_{\mp}^{2}}{\psi_{kw}}\cos\theta_{w}-\frac{k_{z}}{k}\right)\Theta_{\pm}\right]^{1/2}.\label{eq: MHD mid k}\end{equation}
Here $v_{+}=v_{F}/c_{s}$ and $v_{-}=v_{S}/c_{s}$ are the normalized fast and  slow phase velocities, $\tilde{\zeta_{w}}=1+\zeta_{w}$, and
\begin{equation}
\Theta_{\pm}\equiv\frac{k}{k_{z}}\frac{v_{\pm}^{3}(1-v_{\mp}^{2})}{(1-v_{\mp}^{2})^{2}+v_{\pm}^{2}(1-2v_{\mp}^{2})+\frac{k_{\perp}^{2}v_{A}^{2}}{k^{2}c_{s}^{2}}+\frac{k^{2}}{k_{z}^{2}}v_{\mp}^{4}v_{\pm}^{2}},\end{equation}
with ${v}_{A}\equiv B_{0}/\!\sqrt{4\pi\langle\rho\rangle}$. The high-$k$  form [Eq.~\eqref{eq: dust eval pert 3}] of the magnetosonic RDI  is similar, but we omit it here (the Alfv\'en wave is also destabilized at high $k$; see \citealt{Hopkins:2018mhdrdi}). 
In Fig.~\ref{fig: MHD}, we compare these analytic results to numerical solutions of the full $10^{\mathrm{th}}$-order grain-fluid dispersion relation, for a variety of magnetosonic resonances at different  angles.


 \begin{figure}
\begin{center}
\includegraphics[width=1.0\columnwidth]{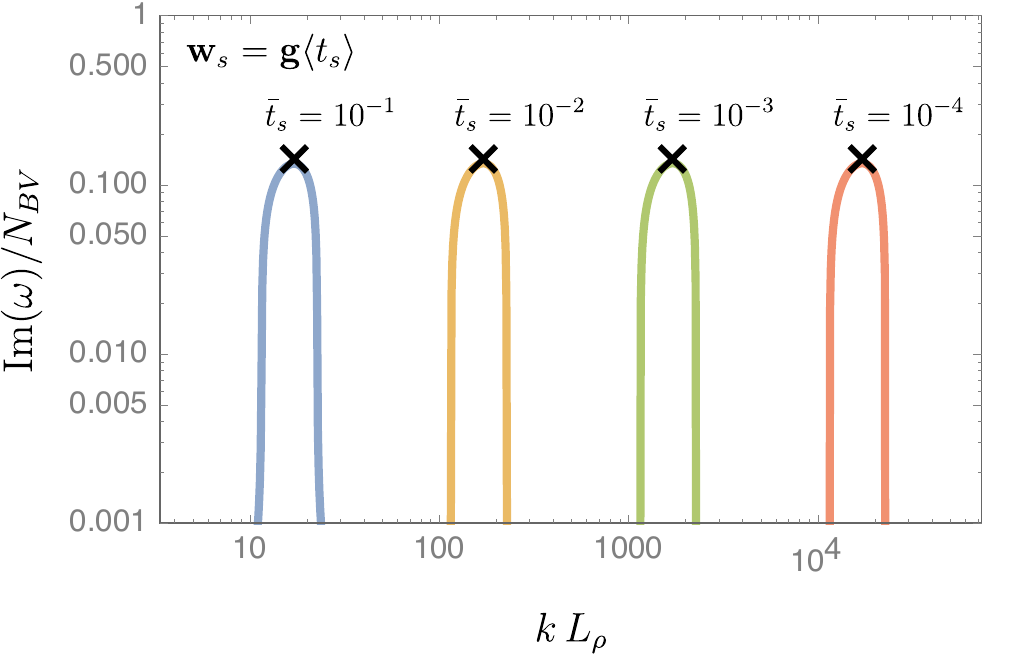}
\caption{\emph{Brunt-V\"ais\"al\"a RDI in a stratified fluid:} solid lines show  numerically calculated 
growth rates for differently sized grains, specified by the normalized stopping time $\bar{t}_{s}\equiv\langle t_{s}\rangle N_{BV}$. We set $\driftvel$ to  the ``natural'' settling of grains due to gravity, $\driftvel=\bm{g}\langle t_{s}\rangle$ ($\theta_{w}=0$), assume Epstein drag \eqref{eq: epstein drag} with $\mu=0.1$, and set $k_{z}=k/2$, $k_{\perp}=\sqrt{3/4}\,k$.    The black crosses show the RDI \eqref{eq: Bouss low k} at resonance, $k=k_{\mathrm{res}}$. 
Smaller grains excite smaller-scale oscillations because they settle more slowly ($V_{\mathrm{wave}}\propto N_{BV}/k\propto \driftvelmag$), but $\mathrm{Im}(\omega)$ is independent of $\langle t_{s}\rangle$  when $\driftvelmag\propto\langle t_{s}\rangle$. 
Because grains move through the atmosphere over timescale $t_{\mathrm{settle}}\sim L_{\rho}/\driftvelmag$, the RDI grows
sufficiently fast to clump grains  (as observed in \citealt{2016A&A...591A.133L}) if $\mathrm{Im}(\omega)/N_{BV}\gtrsim\bar{t}_{s}=\langle t_{s}\rangle N_{BV}$.
}
\label{fig: Boussinesq}
\end{center}
\end{figure}

\subsection{Stratified  fluid}
Our final example is a stratified adiabatic fluid, within the Boussinesq approximation. This instability, in particular its application to planetesimal formation in disks, is treated in detail in \citet{Squire:2017rdi}.
With background gas stratification $\nabla\ln (p_{0}\rho_{0}^{-5/3})=-(5/3)L_{\rho}^{-1}\hat{\bm{z}}$ and gravitational force $\bm{g}=g\hat{\bm{z}}=\tilde{g}\hat{\bm{z}}+\mathcal{O}(\mu)$ (where $\tilde{g}\equiv\rho_{0}^{-1}dp_{0}/dz$), the linearized fluid equations for perturbations $\delta\bm{u}$, $\delta\bar{\rho}=\delta\rho/\langle\rho\rangle$, and $\delta\bar{T}=\delta T/\langle T \rangle$ (temperature) are \citep{1967ApJ...150..571G,1995ApJ...453..380B},
\begin{gather}\partial_{t}\delta\bar{\rho}+L_{\rho}^{-1}\delta u_{z}=0,\quad\delta\bar{\rho}+\delta\bar{T}=0,\nonumber\\
\partial_{t}\delta\bm{u}=-\langle\rho\rangle^{-1}\nabla\delta p+\tilde{g}\,\delta\bar{\rho}\,\hat{\bm{z}},\quad\nabla\cdot\delta\bm{u}=0,\label{eq:BV equations}\end{gather}
where  $\delta p$ enforces $\nabla\cdot\delta\bm{u}=0$.
The system supports oscillations at $\omega_{0}=\pm(k_{\perp}/k)N_{BV}$,  where $N_{BV}=\sqrt{\tilde{g}/L_{\rho}}$ is the Brunt-V\"ais\"al\"a frequency  and $k_{\perp}^{2}=k_{x}^{2}+k_{y}^{2}$.
We set $\driftvelhat=\sin\theta_{w}\hat{\bm{x}}+\cos\theta_{w}\hat{\bm{z}}$, and
resonance occurs when $k_{\mathrm{res}}\driftvelmag\,\psi_{kw}=(k_{\perp}/k)\,N_{BV}$. There is now only one
$k_{\mathrm{res}}$ (for $\driftvelmag$ and $\hat{\bm{k}}$ given),  because $V_{\mathrm{wave}}\propto N_{BV}/k$.
We assume Epstein drag \eqref{eq: epstein drag}, which---using $\delta\rho+\delta T=0$ and $\driftvelmag\ll c_{s}$---implies $\delta t_{s}/\langle t_{s}\rangle\approx-\delta\bar{\rho}/2$.

Inserting Eq.~\eqref{eq:BV equations} and $k=k_{\mathrm{res}}$ into Eq.~\eqref{eq: dust eval pert}, we obtain the ``Brunt-V\"ais\"al\"a RDI,''
\begin{equation}\omega\approx\frac{k_{\perp}}{k}N_{BV}+i\,{\mu^{1/2}}N_{BV}\left[\frac{\driftvelmag}{4\tilde{g}\langle t_{s}\rangle}\left(\cos\theta_{w}-\frac{k_{z}}{k}\psi_{kw}\right)\right]^{1/2}\label{eq: Bouss low k}\end{equation} 
[the high-$k$ scaling \eqref{eq: dust eval pert 3} is never physically applicable]. Evidently, the RDI is unstable, around $k_{\mathrm{res}}=k_{\perp}N/\driftvel\cdot\bm{k}$, unless grains stream exactly against gravity ($\theta_{w}=\pi$ if $L_{\rho}>0$). In Fig.~\ref{fig: Boussinesq}, we compare numerical solutions with Eq.~\eqref{eq: Bouss low k} for gravitationally settling grains ($\driftvel=\bm{g}\langle t_{s}\rangle$), showing the agreement at $k=k_{\mathrm{res}}$.  A compressible treatment reveals minor corrections to Eq.~\eqref{eq: Bouss low k} from corrections to the Boussinesq approximation\footnote{The local treatment also requires $\mu^{1/2}\gg (kL_{\rho})^{-1}$, so that the effect of grains is larger than corrections to the gas modes.}; see \citet{Squire:2017rdi}.

\emph{Discussion.---}We have shown that dust grains streaming (with  velocity $\driftvel$) through a
fluid    are usually  unstable. Specifically, a ``resonant drag instability'' (RDI) occurs \emph{whenever the dust streaming frequency $\bm{k}\cdot\driftvel$ matches the frequency of a fluid wave} $\omega_{0}(\bm{k})$, except for pathological forms of the dust-to-fluid coupling  [see Eqs.~\eqref{eq: dust eval pert}--\eqref{eq: dust eval pert 3}].
All RDIs generically cause  grains to clump spatially as they grow, and will also seed turbulence if sufficiently strong. This could have potentially important consequences for a wide variety of 
astrophysical regions and processes, including planetesimal formation, cool-star winds, AGN torii and winds, starburst regions, HII regions, supernova ejecta, and the circumgalactic medium. Extended discussion of these implications follows in \citet{Hopkins:2017rdi,Hopkins:2018mhdrdi,Squire:2017rdi}.

Rather than exploring any one family of RDIs in detail, our purpose here has been to demonstrate the existence of RDIs and provide an algorithm for identifying its different variants. However, for the sake of illustration, we have provided several examples of the RDI in different fluid systems. In hydrodynamics and MHD, 
the constant phase velocity of linear waves ($\omega_{0}\propto k$) implies that RDI growth rates increase indefinitely as $k\rightarrow\infty$, in the absence of viscosity or resistivity. In MHD, slow waves are destabilized for {any}  $\driftvelmag=|\driftvel|$ (Fig.~\ref{fig: MHD}); in hydrodynamics,  sound waves are destabilized  whenever $\driftvelmag>c_{s}$ (Fig.~\ref{fig: hydro}).
Our final example---a stratified fluid---illustrates the RDI with Brunt-V\"ais\"al\"a oscillations and shows that small grains settling through a stratified atmosphere are unstable. Extensions to other systems (e.g., other fluids or charged grains) are straightforward, given the simplicity of the perturbed eigenvalues \eqref{eq: general eval pert}. For example, as shown in \citet{Squire:2017rdi},  the maximum growth rate of the well-known  disk ``streaming instability''  \citep{2005ApJ...620..459Y} at $\mu<1$ can  be  calculated as the ``epicyclic RDI'' using Eq.~\eqref{eq: dust eval pert}.

Let us finish by reiterating the algorithm presented here for finding drag-induced instabilities in dust-laden fluids: match $\bm{k}\cdot \driftvel$ to an oscillation mode (wave) of the 
fluid; Eq.~\eqref{eq: dust eval pert} or \eqref{eq: dust eval pert 3} then says that the system is most likely unstable, and 
gives the growth rate of the resonant drag modes.

\acknowledgments
We thank A.~R.~Bell, J.~W.~Burby, E. Quataert, and E.~S.~Phinney for  enlightening discussion. 
Support for JS \& PFH was provided by an Alfred P. Sloan Research Fellowship, NASA ATP Grant NNX14AH35G, and NSF Collaborative Research Grant \#1411920 and CAREER grant \#1455342. JS was funded in part by the Gordon and Betty Moore Foundation through Grant GBMF5076 to Lars Bildsten, Eliot Quataert and E. Sterl Phinney.


\end{document}